%% file: final.tex
\documentclass{article}
\usepackage{amsmath, amsfonts, amsthm, amssymb}

\usepackage{times}
\usepackage{graphicx}
\usepackage{subfigure}

\usepackage{natbib}

\usepackage{algorithm}
\usepackage{algorithmic}

\usepackage{amsmath,amssymb}

\usepackage{hyperref}

\usepackage[accepted]{icml2014}

\renewcommand{\qed}{\quad \ensuremath{\blacksquare}}    
\newcommand{\inv}{^{-1}}                            
\newcommand{\sminus}{\backslash}                    
\newcommand{\N}{\mathbb{N}}                         
\newcommand{\R}{\mathbb{R}}                         
\newcommand{\Se}{\mathcal{S}}                       
\newcommand{\e}{\varepsilon}                        
\newcommand{\X}{\mathcal{X}}                        
\newcommand{\E}{\mathbb{E}}                         
\newcommand{\V}{\mathbb{V}}                         
\newcommand{\pr}{\mathbb{P}}                        
\newcommand{\cpest}{\widehat{p}_h}                  
\newcommand{\cqest}{\widehat{q}_h}                  
\newcommand{\pest}{\widetilde{p}_h}                 
\newcommand{\qest}{\widetilde{q}_h}                 
\newcommand{\vx}{\vec{x}}                           
\newcommand{\vy}{\vec{y}}                           
\newcommand{\vv}{\vec{v}}                           
\newcommand{\vu}{\vec{u}}                           
\newcommand{\vi}{{\vec{i}}}                         
\newcommand{\dist}{\operatorname{dist}}             
\newcommand{\B}{\mathcal{B}}                        
\newcommand{\acro}[1]{\textsc{\MakeLowercase{#1}}}

\icmltitlerunning{Exponential Concentration for Divergence Estimation}

\usepackage{natbib}
\usepackage[disable]{todonotes}

\begin{document}
\twocolumn[
\icmltitle{Generalized Exponential Concentration Inequality\\for R\'enyi
Divergence Estimation}

\icmlauthor{Shashank Singh}{sss1@andrew.cmu.edu}
\icmladdress{Carnegie Mellon University,
            5000 Forbes Ave., Pittsburgh, PA 15213 USA}
\icmlauthor{Barnab\'{a}s P\'{o}czos}{bapoczos@cs.cmu.edu}
\icmladdress{Carnegie Mellon University,
            5000 Forbes Ave., Pittsburgh, PA 15213 USA}

\icmlkeywords{divergence estimation, renyi divergence, finite sample bound,
exponential concentration bound}
\vskip 0.3in
]

\input{AbstractIntro}

\section{Problem and Assumptions} \label{sec:Problem}

\subsection{Notation}
We use the notation of multi-indices common in multivariable calculus to index
several expressions. For example, for analytic functions $f : \R^d \to \R$,
\[f(\vy) = \sum_{\vi \in \N^d} \frac{D^\vi f(\vx)}{\vi!} (\vy - \vx)^\vi,\]
where $\N^d$ is the set of $d$-tuples of natural numbers,

\[\vi! := \prod_{k = 1}^d i_k!, \quad \quad
    (\vy - \vx)^\vi := \prod_{k = 1}^d (y_k - x_k)^{i_k}\]
and
\[
    D^\vi f := \frac{\partial^{|\vi|} f}
                   {\partial^{i_1}x_1\cdots\partial^{\alpha_d}x_d}, \quad \quad
    \mbox{for} \quad \quad
    |\vi| := \sum_{k = 1}^d i_k.
\]
We also use the multinomial theorem, which states that,
$\forall k \in \N, \vx \in \R^d$.
\begin{equation}
\label{eq:multinom}
\left( \sum_{i = 1}^d x_i \right)^k
    = \sum_{|\vi| = k} \frac{k!}{\vi!} \vx^\vi.
\end{equation}

\subsection{Problem}
For a given $d \geq 1$, consider random $d$-dimensional real vectors $X$ and
$Y$ in the unit cube $\X := [0,1]^d$, distributed according to densities
$p,q : \X \to \R$, respectively. For a given
$\alpha \in (0,1) \cup (1,\infty)$, we are interested in using a random sample
of $n$ i.i.d. points from $p$ and $n$ i.i.d. points from $q$ to estimate the
R\'enyi $\alpha$-divergence
\[D_\alpha(p\|q)
    = \frac{1}{\alpha - 1}
        \log \left(
          \int_\X
            p^\alpha(\vx)q^{1 - \alpha}(\vx)
          \, d\vx
        \right).
\]

\subsection{Assumptions}
\label{sec:Assumptions}
{\bf Density Assumptions:}\\
We assume that $p$ and $q$ are in the bounded H\"{o}lder class
$\Sigma_\kappa(\beta,L,r)$. That is, for some $\beta \in (0,\infty)$,
if $\ell = \lfloor \beta \rfloor$ is the greatest integer with
$\ell < \beta$, $\forall \vi \in \N^d$ with $|\vi| = \ell$,
the densities $p$ and $q$ each satisfy a $\beta$-H\"{o}lder condition in the
$r$-norm ($r \in [1,\infty)$):
\begin{align}
\notag
|D^\vi p(\vx + \vv) - D^\vi p(\vx)|
 &  \leq L\|\vv\|_r^{\beta - \ell}    \\
\label{ineq:holder_cond}
 &  = L\left( \sum_{i = 1}^d |v_i|^r \right)^{(\beta-\ell)/r},
\end{align}
and, furthermore, there exist $\kappa = (\kappa_1,\kappa_2) \in (0,\infty)^2$
with $\kappa_1 \leq p,q \leq \kappa_2$. We could take $p$ and $q$ to be in
different H\"{o}lder classes $\Sigma_{\kappa_p}(\beta_p,L_p,r_p)$
and $\Sigma_{\kappa_q}(\beta_q,L_q,r_q)$, but the bounds we show depend,
asymptotically, only on the weaker of the conditions on $p$ and $q$ (i.e.,
$\min\{\beta_p,\beta_q\}, \max\{L_p,L_q\}$, etc.).

It is worth commenting on the case that $p$ (similarly, $q$) is $\gamma$ times
continuously differentiable for a positive integer $\gamma$. Since $\X$ is
compact, the $\gamma$-order derivatives of $p$ are bounded. Hence, since $\X$
is convex, the $(\gamma - 1)$-order derivatives of $p$ are Lipschitz, by the
Mean Value Theorem. Consequently, any degree of continuous differentiability
suffices for this assumption.

The existence of an upper bound $\kappa_2$ is trivial, since $p,q$ are
continuous and $\X$ is compact. The existence of a positive lower bound
$\kappa_1$ for $q$ is natural, as otherwise R\'enyi-$\alpha$ divergence
may be $\infty$. The existence of $\kappa_1$ for $p$ is a technical necessity
due to certain singularities at $0$ (see the Logarithm Bound in Section 6.1).
However, in the important special case of R\'enyi-$\alpha$ entropy (i.e.,
$q$ is the uniform distribution), the assumption of $\kappa_1$ for $p$ can be
dropped via an argument using Jensen's Inequality.

As explained later, we also desire $p$ and $q$ to be nearly constant near the
boundary
$\partial \X = \{\vx \in \X : x_j \in \{0,1\} \mbox{ for some } j \in [d]\}$ of
$\X$. Thus, we assume that, for any sequence
$\{\vx_n\}_{n = 1}^\infty \in \X$ with $\dist(\vx_n,\partial \X) \to 0$ as
$n \to \infty$, $\forall \vi \in \N^d$ with $1 \leq |\vi| \leq \ell$,
\[
\lim_{n \to \infty} D^\vi q(\vx_n) = \lim_{n \to \infty} D^\vi p(\vx_n) = 0.
\]

{\bf Kernel Assumptions:}\\
We assume the kernel $K : \R \to \R$ has bounded support
$[-1,1]$ and the following properties (with respect to the Lebesgue measure):
\[\int_{-1}^1 K(u) \, du = 1, \quad \int_{-1}^1 u^jK(u) \, du = 0,
\]
for all $j \in \{1,\dots,\ell\}$, and $K$ has finite $1$-norm, i.e.,
\begin{equation}
\label{eq:k_int}
    \int_{-1}^1 |K(u)| \, du = \|K\|_1 < \infty.
\end{equation}

\section{Estimator} \label{sec:estimator}
\begin{figure}[h!]
\begin{center}
\includegraphics[width=0.4\textwidth,natwidth=610,natheight=642]{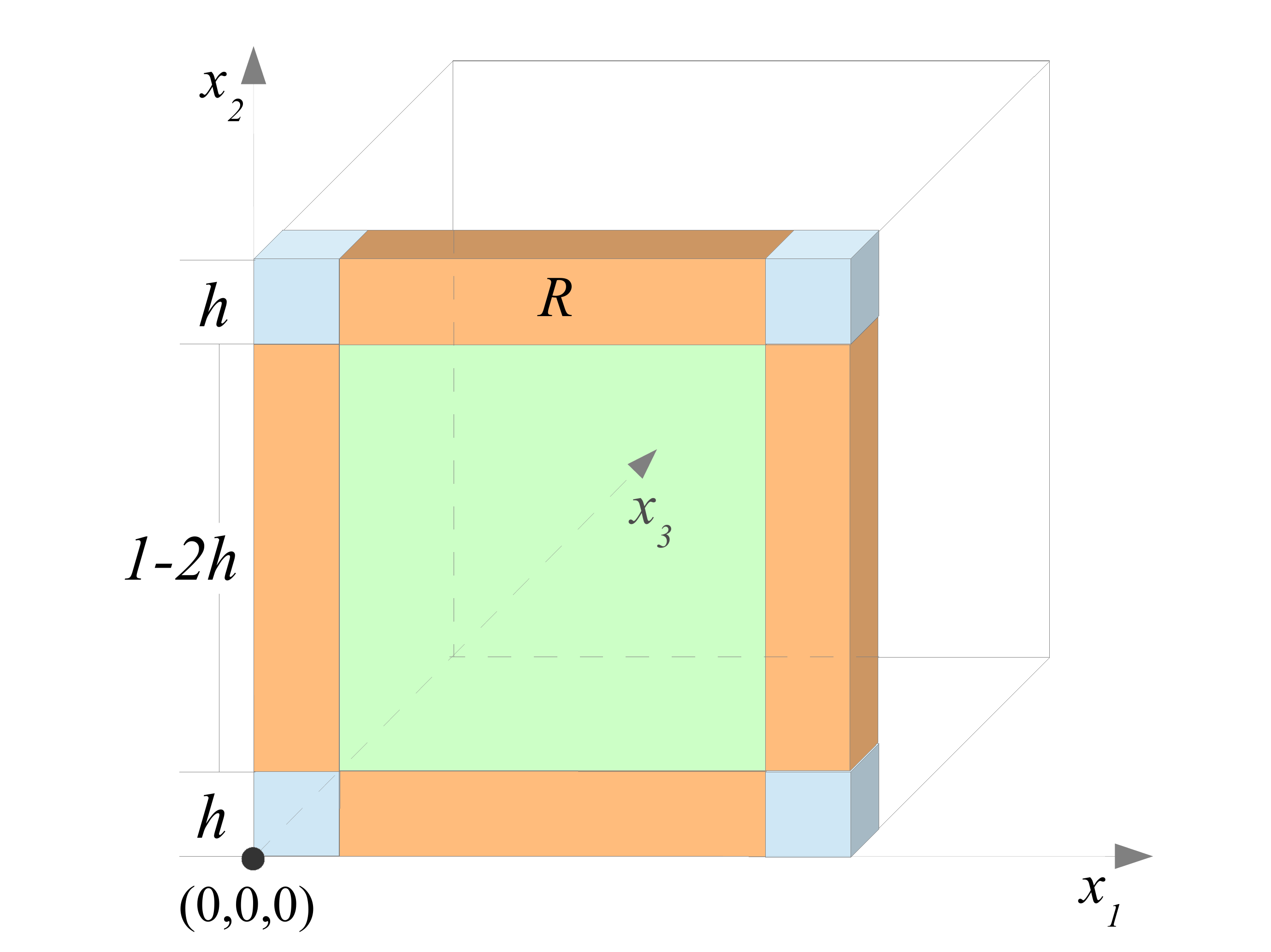}
\end{center}
\vspace{-3mm}
\caption{Illustration of regions $C_{(S_1,S_2,S_3)}$ with $3 \in S_1$. The
region labeled $R$ corresponds to $S_1 = \{3\},S_2 = \{1\},S_3 = \{2\}$.}
\label{fig:cube}
\end{figure}

Let $[d] := \{1,2,\ldots,d\}$, and let
\begin{align*}
\Se := \left\{ (S_1,S_2,S_3) \right. : \; & S_1 \cup S_2 \cup S_3 = [d], \\
& \left. S_i \cap S_j = \emptyset \mbox{ for } i \neq j \right\}
\end{align*}
denote the set of partitions of $[d]$ into $3$ distinguishable parts. For a
bandwidth $h \in (0,1)$ (to be specified later), for each $S \in \Se$, define
the region
\begin{align*}
C_S = \left\{x \in \X : \right.
      &   \;\forall i \in S_1, 0 \leq x_i \leq h, \\
      &   \;\forall j \in S_2, h < x_j < 1 - h,   \\
      &   \left. \forall k \in S_3, 1 - h \leq x_k \leq 1 \right\}
\end{align*}
and the regional kernel $K_S : [-1,2]^d \times \X \to \R$ by
\begin{align*}
K_S(\vx,\vy) :=
 &          \prod_{j \in S_1} K \left(\frac{x_j + y_j}{h}\right)
    \cdot   \prod_{j \in S_2} K \left(\frac{x_j - y_j}{h}\right)  \\
 &  \cdot   \prod_{j \in S_3} K \left(\frac{x_j - 2 + y_j}{h}\right).
\end{align*}
Note that $\{C_S : S \in \Se\}$ partitions $\X$ (as illustrated in
Figure~\ref{fig:cube}), up to intersections of measure zero, and that $K_S$ is
supported only on $[-1,2]^d \times C_S$. The term
$K \left( \frac{x_j + y_j}{h} \right)$ corresponds to reflecting $y$ across the
hyperplane $x_j = 0$, whereas the term $K \left(\frac{x_j - 2 + y_j}{h}\right)$
reflects $y$ across $x_j = 1$, so that $K_S(\vx,y)$ is the product kernel (in
$x$), with uniform bandwidth $h$, centered around a reflected copy of $y$.

We define the ``mirror image'' kernel density estimator
\[\pest(\vx) = \frac{1}{nh^d} \sum_{i = 1}^n \sum_{S\in \Se} K_S(\vx,\vx^i),\]
where $\vx^i$ denotes the $i^{th}$ sample.
Since the derivatives of $p$ and $q$ vanish near $\partial \X$, $p$ and $q$ are
approximately constant near $\partial \X$, and so the mirror image estimator
attempts to reduce boundary bias by mirroring data across $\partial \X$ before
kernel-smoothing. We then clip the estimator at $\kappa_1$ and $\kappa_2$:
\[\cpest(x) = \min(\kappa_2,\max(\kappa_1,\pest(x))).\]

Finally, we plug our clipped density estimate into the following plug-in
estimator for R\'enyi $\alpha$-divergence:
\begin{align}
\label{def:pluginest}
D_\alpha(p \| q)
 &  := \frac{1}{\alpha - 1}
        \log\left( \int_\X p^\alpha(\vx)q^{1 - \alpha}(\vx) \, d\vx \right)
        \notag \\
 &  = \frac{1}{\alpha - 1}
        \log\left( \int_\X f(p(\vx),q(\vx)) \, d\vx \right)
\end{align}
for $f : [\kappa_1,\kappa_2]^2 \to \R$ defined by
$f(x_1,x_2) := x_1^\alpha x_2^{1 - \alpha}$.
Our $\alpha$-divergence estimate is then $D_\alpha(\cpest \| \cqest)$.

\section{Main Result} \label{sec:main-result}
Rather than the usual decomposition of mean squared error into squared bias and
variance, we decompose the error $|D_\alpha(\cpest\|\cqest)~-~D_\alpha(p\| q)|$
of our estimatator into a bias term and a variance-like term via the triangle
inequality:
\begin{align*}
|D_\alpha(\cpest\|\cqest) - D_\alpha(p\| q)|
 &  \leq \underbrace{|D_\alpha(\cpest\|\cqest) - \E D_\alpha(\cpest\|\cqest)|}
        _{\text{variance-like term}}\\
 &  + \underbrace{|\E D_\alpha(\cpest\|\cqest) - D_\alpha(p\| q)|}
        _{\text{bias term}}.
\end{align*}
We will prove the ``variance'' bound
\[\pr \left( |D_\alpha(\cpest,\cqest) - \E D_\alpha(\cpest,\cqest)| > \e \right)
    \leq 2\exp \left( -\frac{k_1\e^2n}{\|K\|_1^{2d}} \right),
\]
and the bias bound
\[|\E D_\alpha(\cpest\|\cqest) - D_\alpha(p\| q)|
    \leq k_2 \left(h^\beta + h^{2\beta} + \frac{1}{nh^d}\right),
\]
where $k_1,k_2$ are constant in the sample size $n$ and bandwidth $h$ (see
(\ref{ineq:bias_consts}) and (\ref{ineq:var_consts}) for exact values of these
constants). The variance bound does not depend on $h$, while the bias bound is
minimized by $h \asymp n^{-\frac{1}{d + \beta}}$, giving the convergence rate
\[|\E D_\alpha(\cpest \| \cqest) - D_\alpha(p \| q)|
    \in O \left( n^{-\frac{\beta}{d + \beta}} \right).\]

Note that we can use this exponential concentration bound to obtain a bound on
the variance of $D(\cpest\|\cqest)$. If $F~:~[0,\infty)\to\R$ is the
cummulative distribution of the squared deviation of $D_\alpha(\cpest\|\cqest)$
from its mean,
\begin{align*}
1 - F(\e)
 &  = \pr \left( (D_\alpha(\cpest,\cqest) - \E D_\alpha(\cpest,\cqest))^2
                                                            > \e \right)    \\
 &  \leq 2\exp \left( -\frac{k_1n}{\|K\|_1^{2d}} \right).
\end{align*}
Thus,
\begin{align}
\notag
\V[D_\alpha(\cpest\|\cqest)]
 &  = \E\left[ D_\alpha(\cpest,\cqest) - \E D_\alpha(\cpest,\cqest)\right]^2 \\
\notag
 &  = \int_0^\infty \left( 1 - F(\e) \right) \, d\e    \\
\notag
 &  \leq \int_0^\infty 2\exp \left(-\frac{k_1n\e}{\|K\|_1^{2d}}\right) \, d\e \\
\label{ineq:fin_var_bnd}
 &  = 2 \frac{\|K\|_1^{2d}}{k_1}n\inv.
\end{align}
We then have a mean squared-error of
\[\E \left[ \left( D(\cpest\|\cqest) - D(p\|q) \right)^2 \right]
    \in O \left( n\inv + n^{-\frac{2\beta}{d + \beta}} \right).
\]
which is in $O(n\inv)$ if $\beta \geq d$ and in
$O\left( n^{-\frac{2\beta}{d + \beta}} \right)$ otherwise. This asymptotic rate
is consistent with previous bounds in density functional estimation
\citep{birge95estimation,2010arXiv1012.4188S}.

\section{Proof of Main Result} \label{sec:Proofs}
\subsection{Lemmas}

{\bf Bound on Derivatives of $f$:}
Let $f$ be as in (\ref{def:pluginest}). Since $f$ is analytic on the compact
domain $[\kappa_1,\kappa_2]^2$, there is a constant $C_f \in \R$, depending
only on $\kappa$, and $\alpha$, such that,
$\forall \xi \in (\kappa_1,\kappa_2)^2$,
\begin{align}
\notag
&  \left| \frac{\partial f}{\partial x_1} (\xi) \right|,
   \left| \frac{\partial f}{\partial x_2} (\xi) \right|,
   \left| \frac{\partial^2 f}{\partial x_1^2} (\xi) \right|,   \\
\label{ineq:parts}
&   \left| \frac{\partial^2 f}{\partial x_2^2} (\xi) \right|,
    \left| \frac{\partial^2 f}{\partial x_1x_2} (\xi) \right|
 \leq C_f.
\end{align}
$C_f$ can be computed explicitly by differentiating $f$ and observing that the
derivatives of $f$ are monotone in each argument. We will use this bound later
in conjunction with the Mean Value and Taylor's theorems.

{\bf Logarithm Bound:} If $g,\hat g : \X \to \R$ with $0 < c \leq g,\hat g$ for
some $c \in \R$ depending only on $\kappa$ and $\alpha$, then, by the Mean
Value Theorem, there exists $C_L$ depending only on $\kappa$ and $\alpha$ such
that
\begin{align}
\notag
 & \left| \log \left( \int_\X \hat g(\vx) \, d\vx \right)
        - \log \left( \int_\X g(\vx) \, d\vx \right) \right|        \\
\label{ineq:log}
 & \leq C_L \int_\X \left| \hat g(\vx) - g(\vx) \right| \, d\vx.
\end{align}
We will use this bound to eliminate logarithms from our calculations.

{\bf Bounds on Derivatives of $p$:}
Combining the assumption that the derivatives of $p$ vanish on $\partial \X$
and the H\"{o}lder condition on $p$, we bound the derivatives of $p$
\emph{near} $\partial \X$. In particular, we show that, if $\vi \in \N^d$
has $1 \leq |\vi| \leq \ell$, then,
$\forall \vx \in \B := \{\vx \in \X : \dist(\vx,\partial \X) \leq h\}$
\begin{equation}
\label{ineq:p_boundary_parts}
|D^\vi p(\vx)|
    \leq \frac{Lh^{\beta - |\vi|}}{(\ell - |\vi|)!}.
\end{equation}
\emph{Proof:} We proceed by induction on $|\vi|$, as $|\vi|$ decreases from
$\ell$ to $0$. The case $|\vi| = \ell$ is precisely the H\"{o}lder assumption
(\ref{ineq:holder_cond}). Now suppose that we have the desired bound for
derivatives of order $|\vi| + 1$. Let
$\vx \in \partial \X$, $\vu = (0,\dots,0,\pm1,0,\dots,0) \in \R^d$, where
$u_j = \pm1$. If $\vy + h\vu \in \X$ (any $\vx \in \B$ is clearly of this form,
for some $j \in [d]$), then
\begin{align*}
|D^\vi p(\vy + \vu)|
 & \leq \int_0^h \left| \frac{\partial}{\partial x_j}
                    D^\vi p(\vy + t\vu) \right| \, dt            \\
 & \leq \int_0^h \frac{Lt^{\beta - (|\vi| + 1)}}{(\ell - |\vi| - 1)!} \, dt    \\
 & = \frac{Lh^{\beta - |\vi|}}{(\beta - |\vi|)(\ell - |\vi| - 1)!}
   \leq \frac{Lh^{\beta - |\vi|}}{(\ell - |\vi|)!}.
\end{align*}
The desired result follows by induction on $|\vi|$. \qed

{\bf Bound on Integral of Mirrored Kernel:}
A key property of the mirrored kernel is that the mass of the kernel over $\X$
is preserved, even near the boundary of $\X$, as the kernels about the
reflected data points account exactly for the mass of the kernel about the
original data point that is not in $\X$. In particular, $\forall \vy \in \X$,
\begin{equation}
\label{eq:mkmass}
\sum_{S \in \Se} \int_\X |K_S(\vx,\vy)| \, d\vx = h^d\|K\|_1^d.
\end{equation}
\begin{figure}[h!]
\begin{center}
\includegraphics[width=0.4\textwidth,natwidth=610,natheight=642]{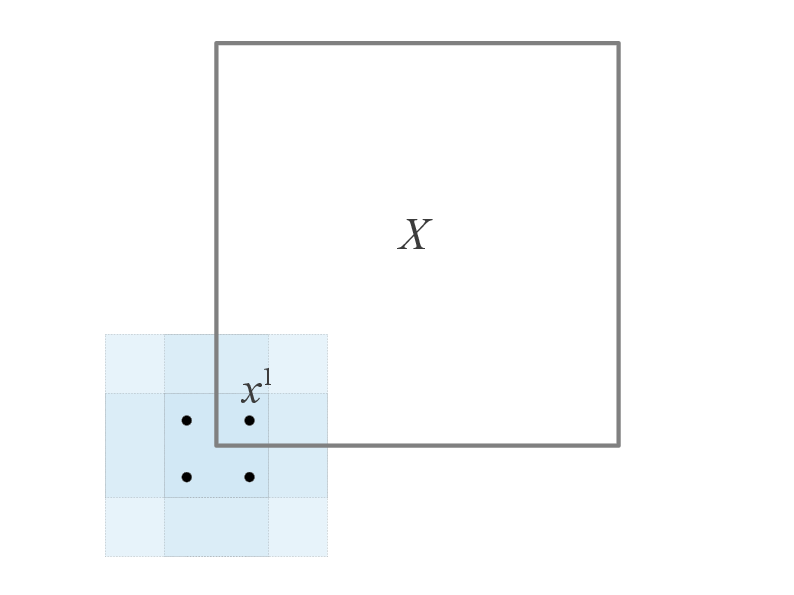}
\end{center}
\vspace{-3mm}
\caption{A data point
$x^1 \in C_{(\{1,2\},\emptyset,\emptyset)} \subset [0,1]^2$, along with its
three reflected copies. The sum of the integrals over $\X$ of the four kernels
(shaded) is $\|K\|_1$.}
\label{fig:mirror}
\end{figure}
\emph{Proof:}
For each $S \in \Se$, the change of variables
\begin{align*}
& u_j = -x_j,     & j \in S_1 \\
& u_j =  x_j,     & j \in S_2 \\
& u_j = 2 - x_j,  & j \in S_3
\end{align*}
returns the reflected data point created by $K_S$ back onto its original data
point. Applying this change of variables gives
\begin{align*}
 \sum_{S \in \Se} \int_\X |K_S(\vx,\vy)| \, d\vx
 &  = \int_{[-1,2]^d} \left|K^d\left(\frac{u - y}{h}\right)\right| \, d\vx,
\end{align*}
where $\displaystyle K^d(\vx)~:=~\prod_{i = 1}^d K(x_i)$ denotes the product
kernel. Rescaling, translating, and applying Fubini's Theorem,
\begin{align*}
 &  \sum_{S \in \Se} \int_\X |K_S(\vx,\vy)| \, d\vx
    = h^d\int_{[-1,1]^d} |K^d(\vx)| \, d\vx         \\
 &  = h^d\left( \int_{-1}^1 |K(u)| \, du \right)^d
    = h^d\|K\|_1^d. \qed
\end{align*}

\subsection{Bias Bound}
The following lemma bounds the integrated square bias of $\pest$ for an
arbitrary $p \in \Sigma_{\kappa_1,\kappa_2}(\beta,L,r)$. We write the bias of
$\pest$ at $\vx \in \X$ as $B_p(\vx) = \E\pest(\vx) - p(\vx)$.

{\bf Bias Lemma:} There exists a constant $C > 0$ such that
\begin{equation}
\label{ineq:bias_lemma}
    \int_\X B_p^2(\vx) \, d\vx \leq Ch^{2\beta}.
\end{equation}

We consider separately the interior $\mathcal{I} := (h,1 - h)^d$ and
boundary $\B$ (noting $\X = \mathcal{I} \cup \mathcal{B}$). By a standard
result for kernel density estimates of H\"{o}lder continuous functions (see,
for example, Proposition 1.2 of \citet{Tsybakov:2008:INE:1522486}),
\[\int_{\mathcal{I}} B_p^2(x) \, d\vx \leq C_2h^{2\beta}.\]
(In particular, this holds for the constant
$\displaystyle C_2 := \frac{L}{\ell!}\|K\|_1^d$.)

We now show that
$\displaystyle \int_{\mathcal{B}} B_p^2(\vx) \, d\vx \leq C_3^2h^{2\beta}$.

Suppose $S = (S_1,S_2,S_3) \in \Se \sminus\{(\emptyset,[d],\emptyset)\}$
(as $C_{(\emptyset,[d],\emptyset)} = \mathcal{I}$). We wish to bound
$|B_p(\vx)|$ on $C_S$. To simplify notation, by geometric symmetry, we
may assume $S_3 = \emptyset$. Let $\vu \in [-1,1]^d$, and define $\vy_S \in \X$
by $(y_S)_i = hu_i - x_i, \forall i \in S_1$ and
$(y_S)_i = x_i - hu_i, \forall i \in S_2$ (this choice arises from the change
of variables we will use in (\ref{eq:CoV})). By the H\"{o}lder
condition (\ref{ineq:holder_cond}) and the choice of $y_S$,
\begin{align*}
 &  \left|p(\vy_S) - \sum_{|\vi| \leq \ell}
     \frac{D^\vi p(x)}{\vi!} (\vy_S - \vx)^\vi \right|
    \leq L\|\vy_S - \vx\|_r^\beta                               \\
 &  = L \left( \sum_{j \in S_1} |2x_j + hu_j|^r
                 + \sum_{j \in S_2} |hu_j|^r \right)^{\beta/r}
\end{align*}
Since each $|u_j| \leq 1$ and, for each $i \in S_1$, $0 \leq x_j \leq h$,
\begin{align*}
 &  \left|p(\vy_S) - \sum_{|\alpha| \leq \ell}
     \frac{D^\vi p(x)}{\vi!} (\vy_S - \vx)^\vi \right|\\
 &  = L \left( \sum_{j \in S_1} (3h)^r
                 + \sum_{j \in S_2} h^r \right)^{\beta/r}\\
 &  \leq L \left(d\left(3h\right)^r\right)^{\beta/r}
    = L \left(3d^{1/r} h\right)^\beta.
\end{align*}
Rewriting this using the triangle inequality
\begin{align}
\notag
 & |p(\vy_S) - p(\vx)|  \\
 & \leq L \left(3d^{1/r} h\right)^\beta
   + \left| \sum_{1 \leq |\alpha| \leq \ell}
        \frac{D^\alpha p(\vx)}{\alpha!} (\vy_S - \vx)^\alpha \right|.
\label{ineq:tribnd}
\end{align}
Observing that $(\vy - \vx)^\vi \leq (3h)^{|\vi|}$ and applying the bound on
the derivatives of $p$ near $\partial \X$ (as computed in
(\ref{ineq:p_boundary_parts})),
\begin{align*}
    \left| \sum_{1 \leq |\vi| \leq \ell}
        \frac{D^\vi p(x)}{\vi!} (\vy_S - \vx)^\vi \right|      \\
 \leq \left| \sum_{|\vi| \leq \ell}
        \frac{Lh^{\beta - |\vi|}}{(\ell - |\vi|)!\vi!} (3h)^{|\vi|} \right|   \\
 = Lh^\beta \sum_{k = 0}^\ell \sum_{|\vi| = k} \frac{3^{|\vi|}}{(\ell - k)!\vi!}    \\
 \leq Lh^\beta \sum_{k = 0}^\ell \frac{1}{k!(\ell - k)!} \sum_{|\vi| = k} \frac{k!3^{|\vi|}}{\vi!}
\end{align*}
Then, applying the multinomial theorem (\ref{eq:multinom}) followed by the
binomial theorem gives
\begin{align*}
 \left| \sum_{1 \leq |\vi| \leq \ell}
        \frac{D^\vi p(x)}{\vi!} (\vy_S - \vx)^\vi \right|
 & \leq Lh^\beta \sum_{k = 0}^\ell \frac{(3d)^k}{k!(\ell - k)!}   \\
 & =    Lh^\beta \frac{1}{\ell!} \sum_{k = 0}^\ell \frac{\ell!}{(\ell - k)!k!}(3d)^k \\
 & =    Lh^\beta \frac{(3d + 1)^\ell}{\ell!}.
\end{align*}
Combining this bound with (\ref{ineq:tribnd}) gives
\begin{align}
\label{ineq:loc_bias_lemma}
|p(\vy_S) - p(\vx)|
    = C_3 h^\beta,
\end{align}
where $C_3$ is the constant (in $n$ and $h$)
\begin{align}
\label{eq:biasC}
C_3 := L \left( \left( 3d^{1/r} \right)^\beta + \frac{(3d + 1)^\ell}{\ell!} \right).
\end{align}
For $x \in C_S$, we have
$\pest(\vx)  = \frac{1}{nh^d} \sum_{i = 1}^n K_S(\vx,\vx^i)$, and thus, by a
change of variables, recalling that $K^d(\vx)$ denotes the product kernel,
\begin{align}
\notag
\E\pest(\vx)
 &  = \frac{1}{h^d} \int_\X K_S(\vx,\vu) p(\vu) \, d\vu    \\
\label{eq:CoV}
 &  = \int_{[-1,1]^d} K^d(\vv) p(\vy_S) \, d\vv,
\end{align}
Since $\displaystyle \int_{[-1,1]} K^d(\vv) \, d\vv = 1$, by the bound in
(\ref{ineq:loc_bias_lemma}),
\begin{align*}
 & |B_p(x)| = |\E\pest(x) - p(x)|   \\
 & = \left| \int_{[-1,1]^d} K^d(\vv) p(\vy_S) \, d\vv
   - \int_{[-1,1]^d} K^d(\vv) p(\vx) \, d\vv \right| \\
 & \leq \int_{[-1,1]^d} K^d(\vv) |p(\vy_S) - p(\vx)| \, d\vv    \\
 & \leq \int_{[-1,1]^d} K^d(\vv) C_3 h^\beta \, d\vv
   = C_3h^\beta.
\end{align*}
Then, $\int_{\mathcal{B}} B_p^2(x)~\,~d\vx~\leq~C_3^2 h^{2\beta}$ (since the
measure of $\mathcal{B}$ is less than $1$), proving the Bias Lemma. \qed

By Taylor's Theorem, $\forall \vx \in \X$, for some $\xi : \X \to \R^2$ on the
line segment between $(\cpest(\vx),\cqest(\vx))$ and $(p(\vx),q(\vx))$,

\begin{align*}
& \left| \E f(\cpest(\vx), \cqest(\vx)) - f(p(\vx),q(\vx)) \right|  \\
    & = \left|
        \E\frac{\partial f}{\partial x_1}(p(\vx),q(\vx))(\cpest(\vx) - p(\vx)) \right.  \\
    & + \frac{\partial f}{\partial x_2}(p(\vx),q(\vx))(\cqest(\vx) - q(\vx))  \\
    & + \frac12 \left[
            \frac{\partial^2 f}{\partial x_1^2}(\xi)(\cpest(\vx) - p(\vx))^2
        \right.
      + \frac{\partial^2 f}{\partial x_2^2}(\xi)(\cqest(\vx) - q(\vx))^2 \\
    & \quad\quad\quad + \left. \left.
            \frac{\partial^2 f}{\partial x_1 \partial x_2}(\xi)
                    (\cpest(\vx) - p(\vx))(\cqest(\vx) - q(\vx))
        \right] \right| \\
    & \leq C_f \left(
        \left| B_p(\vx) \right|
      + \left| B_q(\vx) \right| \right.
      + \E \left[ \cpest(\vx) - p(\vx) \right]^2    \\
    & + \E \left[ \cqest(\vx) - q(\vx) \right]^2
      + \left. \left| B_p(\vx)B_q(\vx)) \right|
    \right),
\end{align*}
where the last line follows from the triangle inequality and
(\ref{ineq:parts}). Thus, using (\ref{ineq:log}),
\begin{align*}
 & |\E D_\alpha(\cpest \| \cqest) - D_\alpha(p \| q)|   \\
 &  = \left| \frac{1}{\alpha - 1} \left(
        \E \log  \int_\X f(\cpest(\vx),\cqest(\vx)) \, d\vx \right. \right. \\
 &  \quad \left. \left. - \log \int_\X f(p(\vx),q(\vx)) \, d\vx
      \right) \right| \\
 &  \leq \frac{C_L}{|\alpha - 1|}
    \int_\X \left| \E f(\cpest(\vx),\qest(\vx)) - f(p(\vx),q(\vx)) \, d\vx
\right| \\
 &  \leq \frac{C_fC_L}{|\alpha - 1|}
    \int_\X \left| B_p(\vx) \right|
    + \left| B_q(\vx) \right|
    + \E \left[ \cpest(\vx) - p(\vx) \right]^2  \\
 &  + \E \left[ \cqest(\vx) - q(\vx) \right]^2
    + \left| B_p(\vx)B_q(\vx) \right| \, d\vx.
\end{align*}
By H\"{o}lder's Inequality, we then have
\begin{align*}
 & |\E D_\alpha(\cpest \| \cqest) - D_\alpha(p \| q)|   \\
 &  \leq \frac{C_fC_L}{|\alpha - 1|\kappa_1} \left(
    \sqrt{\int_\X B_p^2(\vx) \, d\vx} \right.
    + \sqrt{\int_\X B_q^2(\vx) \, d\vx} \\
 &  + \int_\X \E \left[ \cpest(\vx) - p(\vx) \right]^2
    + \E \left[ \cqest(\vx) - q(\vx) \right]^2 \, d\vx       \\
 &  + \left. \sqrt{\int_\X B_p^2(\vx)
            \int_\X B_q^2(\vx) \, d\vx} \right).
\end{align*}
Applying Lemma 3.1 and a standard result in kernel density estimation (see, for
example, Propositions 1.1 and 1.2 of \citet{Tsybakov:2008:INE:1522486}) gives
\begin{align}
\notag
 & |\E D_\alpha(\cpest \| \cqest) - D_\alpha(p \| q)|   \\
\label{ineq:bias_consts}
 &  \leq \left( C_2 + C_3 \right)h^\beta
    + C_2 h^{2\beta}
    + \kappa_2 \frac{\|K\|_1^d}{nh^d} \\
\notag
 & \leq C\left(h^\beta + h^{2\beta} + \frac{1}{nh^d}\right),
\end{align}
for some $C > 0$ not depending on $n$ or $h$. \qed
\subsection{Variance Bound}
Consider i.i.d. samples $\vx^1,\dots,\vx^n\sim~p, \vy^1,\dots,\vy^n\sim~q$. In
anticipation of using McDiarmid's Inequality \citep{McDiarmid1989}, let
$\cpest'(\vx)$ denote our kernel density estimate with the sample $\vx^j$
replaced by $(\vx^j)'$. By (\ref{ineq:log}),
\begin{align*}
 & |D_\alpha(\cpest\|\cqest) - D_\alpha(\pest'\|\qest')|    \\
 &  = \frac{1}{|\alpha - 1|}
        \left| \log\left(\int_\X f(\cpest(\vx), \cqest(\vx)) \, d\vx\right)
                \right. \\
 &  - \left. \log\left(\int_\X f(\cpest'(\vx),\cqest(\vx)) \, d\vx\right)
                \right| \\
 &  \leq \frac{C_L}{|\alpha - 1|} \int_\X
        \left| f(\cpest(\vx),\cqest(\vx)) - f(\cpest'(\vx),\cqest(\vx))
                \right| \, d\vx.
\end{align*}
Then, applying the Mean Value Theorem followed by (\ref{ineq:parts}) gives, for
some  $\xi : \X \to \R^2$ on the line segment between $(\cpest,\cqest)$
and $(p,q)$,
\begin{align*}
 & |D_\alpha(\cpest\|\cqest) - D_\alpha(\pest'\|\qest')|    \\
 &  \leq \frac{C_L}{|\alpha - 1|}
    \int_\X \left| \frac{\partial f}{\partial x_1} (\xi(\vx))
            (\cpest(\vx) - \cpest'(\vx)) \right| \, dx  \\
 &  \leq \frac{C_LC_f}{|\alpha - 1|}
    \int_\X \left| \cpest(\vx) - \cpest'(\vx) \right| \, d\vx.
\end{align*}
Expanding $\cpest$ as per its construction gives
\begin{align*}
 & |D_\alpha(\cpest\|\cqest) - D_\alpha(\pest'\|\qest')|        \\
 &  \leq \frac{C_LC_f}{|\alpha - 1|}
    \int_\X \left| \pest(\vx) - \pest'(\vx) \right| \, d\vx     \\
 &  \leq \frac{C_LC_f}{|\alpha - 1|nh^d} \sum_{S \in \Se}
    \int_\X \left|K_S(x,x^j) - K_S(x,(x^j)') \right| \, d\vx    \\
 &  \leq \frac{2C_LC_f}{|\alpha - 1|nh^d} \sup_{y \in \X} \sum_{S \in \Se}
    \int_\X \left|K_S(x,y) \right| \, d\vx
    = \frac{2C_LC_f}{|\alpha - 1|n} \|K\|_1^d,
\end{align*}
where the last line follows from the triangle inequality and (\ref{eq:mkmass}).
An identical proof holds if we vary some $\vy^i$ rather than $\vx^i$. Thus,
since we have $2n$ independent samples, McDiarmid's Inequality gives the bound,
\[ \pr \left( |D_\alpha(\cpest,\cqest) - \E D_\alpha(\cpest,\cqest)|
        > \e \right)
   \leq 2\exp \left(
            -\frac{C^2\e^2n}{\|K\|_1^{2d}}
        \right),
\]
\begin{equation}
\label{ineq:var_consts}
\mbox{where } \quad
  C = \frac{|\alpha - 1|}{2C_LC_f}
\end{equation}
depends only on $\kappa$ and $\alpha$. \qed

\section{Experiment} \label{sec:numerical}
\begin{figure}[h!]
\begin{center}
\includegraphics[width=0.5\textwidth]{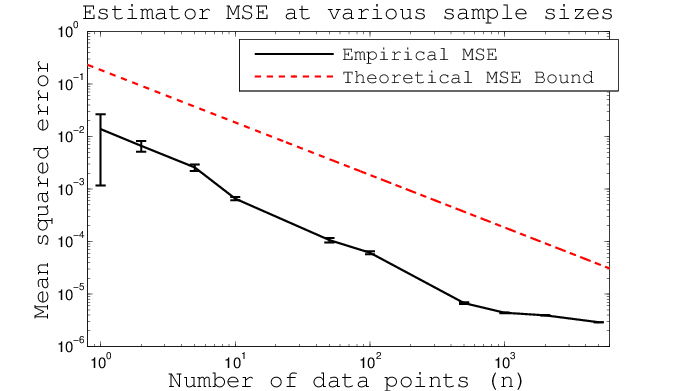}
\end{center}
\vspace{-3mm}
\caption{Log-log plot of mean squared error (computed over 100 trials) of our
estimator for various sample sizes $n$, alongside our theoretical bound. Error
bars indicate standard deviation of estimator over 100 trials.}
\label{fig:MSE}
\end{figure}

We used our estimator to estimate the R\'enyi $\alpha$-divergence between two
normal distributions in $\R^3$ restricted to the unit cube. In particular, for
\[
\vec{\mu}_1 = \begin{bmatrix}
0.3 \\
0.3 \\
0.3
\end{bmatrix},
\vec{\mu}_2 = \begin{bmatrix}
0.7 \\
0.7 \\
0.7
\end{bmatrix},
\Sigma = \begin{bmatrix}
0.2 & 0   & 0   \\
0   & 0.2 & 0   \\
0   & 0   & 0.2 \\
\end{bmatrix},
\]
$p = \mathcal{N}(\vec{\mu}_1,\Sigma),q = \mathcal{N}(\vec{\mu}_2,\Sigma)$.
For each $n \in \{1,2,5,10,50,100,500,1000,2000,5000\}$, $n$ data points were
sampled according to each distribution and constrained (via rejection sampling)
to lie within $[0,1]^3$. Our estimator was computed from these samples, for
$\alpha = 0.8$, using the Epanechnikov Kernel $K(u) = \frac{3}{4} (1 - u^2)$
on $[-1,1]$, with constant bandwidth $h = 0.25$. The true $\alpha$-divergence
was computed directly according to its definition on the (renormalized)
distributions on $[0,1]^3$. The bias and variance of our estimator were then
computed in the usual manner based on 100 trials. Figure~\ref{fig:MSE} shows
the error and variance of our estimator for each $n$.

We also compared our estimator's empirical error to our theoretical bound.
Since the distributions used are infinitely differentiable, $\beta = \infty$,
and so the estimator's mean squared error should converge as $O(n\inv)$. An
appropriate constant multiple was computed from (\ref{ineq:fin_var_bnd}),
(\ref{eq:biasC}), and (\ref{ineq:bias_consts}). The resulting bound is also
shown in Figure~\ref{fig:MSE}.
\vspace{-5mm}
\section{Conclusion} \label{sec:Conclusion}
In this paper we derived a finite sample exponential concentration bound for a
consistent, nonparametric, $d$-dimensional R\'enyi-$\alpha$ divergence
estimator. To the best of our knowledge this is the first such exponential
concentration bound for Renyi divergence.

\bibliography{biblio}
\bibliographystyle{icml2014}

\end{document}

%% file: AbstractIntro.tex
\begin{abstract}
Estimating divergences in a consistent way is of great importance in many
machine learning tasks. Although this is a fundamental problem in nonparametric
statistics, to the best of our knowledge there has been no finite sample
exponential inequality convergence bound derived for any divergence estimators.
The main contribution of our work is to provide such a bound for an estimator
of R\'enyi-$\alpha$ divergence for a smooth H\"older class of densities on the
$d$-dimensional unit cube $[0,1]^d$.
We also illustrate our theoretical results with a numerical experiment.
\end{abstract}

\section{Introduction}
There are several important problems in machine learning and statistics that
require the estimation of the distance or divergence between distributions. In
the past few decades many different kinds of divergences have been defined to
measure the discrepancy between distributions. They include the
Kullback--Leibler (\acro{KL}) \citep{kullback51KL},
R\'enyi-$\alpha$ \citep{renyi61measures,renyi70probability},
Tsallis-$\alpha$ \citep{villmann10mathematical},
Bregman \citep{bregman67divergence},
$L_p$,
maximum mean discrepancy \citep{Borgwardt06MMD},
Csisz\'ar's-$f$ divergence \citep{csiszar67information}
and many others.

Most machine learning algorithms operate on finite dimensional feature vectors.
Using divergence estimators one can develop machine learning algorithms (such
as regression, classification, clustering, and  others) that can operate on
sets and distributions \citep{poczos12kernelimages,oliva13ICML}. Under certain
conditions, divergences can estimate entropy and mutual information. Entropy
estimators are important in
goodness-of-fit testing \citep{goria05new},
parameter estimation in semi-parametric models \citep{Wolsztynski85minimum},
studying fractal random walks \citep{Alemany94fractal},
and texture classification \citep{hero02alpha,hero2002aes}.
Mutual information estimators have been used
in feature selection \citep{peng05feature},
clustering \citep{aghagolzadeh07hierarchical},
optimal experimental design \citep{lewi07realtime}, 
f\acro{MRI} data processing \citep{chai09exploring},
prediction of protein structures \citep{adami04information},
boosting and facial expression recognition \citep{Shan05conditionalmutual}.
Both entropy estimators and mutual information estimators have been used for
independent component and subspace analysis
\citep{radical03,szabo07undercomplete_TCC}, 
as well as for image registration
\citep{kybic06incremental,hero02alpha,hero2002aes}.
For further applications, see
\citet{Leonenko-Pronzato-Savani2008}.

In this paper we will focus on the estimation of R\'enyi-$\alpha$ divergences.
This important class contains the Kullback--Leibler divergence as the
$\alpha\to 1$ limit case and can also be related to the Tsallis-$\alpha$,
Jensen-Shannon, and Hellinger divergences.\footnote{Some of the divergences
mentioned in the paper are distances as well. To simplify the treatment we will
call all of them divergences.}
It can be shown that many information theoretic quantities
(including entropy, conditional entropy, and mutual information)
can be computed as special cases of R\'enyi-$\alpha$ divergence.

In our framework, we assume that the underlying distributions are not given
explicitly. Only two finite, independent and identically distributed (i.i.d.)
samples are given from some unknown, continuous, nonparametric distributions.

Although many of the above mentioned divergences were
defined a couple of decades ago, interestingly there are still many open
questions left to be answered about the properties of their estimators. In
particular, even simple questions, such as rates are unknown for many
estimators, and to the best of our knowledge no finite sample exponential
concentration bounds have ever been derived for divergence estimators.

\textbf{Our main contribution} is to derive an exponential concentration bound
for a particular consistent, nonparametric, R\'enyi-$\alpha$ divergence
estimator. We illustrate the behaviour of the estimator with a numerical
experiment.

\subsection*{Organization}
In the next section we discuss related work (Section~\ref{sec:related}).
In Section~\ref{sec:Problem} we formally
define the R\'enyi-$\alpha$ divergence estimation problem, and
introduce the notation and the assumptions used in the paper.
Section~\ref{sec:estimator} presents the divergence estimator that we study in
the paper.
Section~\ref{sec:main-result} contains our main theoretical contibutions
concerning the exponential concentration bound of the divergence estimator.
Section~\ref{sec:Proofs} contains the proofs of our theorems.
To illustrate the behaviour of the estimator, we provide a simple numerical
experiment in Section~\ref{sec:numerical}.
We draw conclusions in Section~\ref{sec:Conclusion}.

\section{Related Work} \label{sec:related}

Probably the closest work to our contribution is \citet{liu12exponential}, who
derived exponential-concentration bound for an estimator of the two-dimensional
Shannon entropy. We generalize these results in several aspects:
\begin{enumerate}
\item The estimator of \citet{liu12exponential} operates in the unit square $[0,1]^2$. Our
estimator operates on the $d$-dimensional unit hypercube $[0,1]^d$.
\item In \citet{liu12exponential} the exponential concentration inequality was proven for
densities in the H\"{o}lder class $\Sigma_\kappa(2,L,2)$, whereas our inequality
applies for densities in the H\"{o}lder class $\Sigma_\kappa(\beta,L,r)$ for
any fixed $\beta \geq 0, r \geq 1$ (see Section~\ref{sec:Assumptions} for definitions of these
H\"older classes).
\item While \citet{liu12exponential} estimated the Shannon entropy using one i.i.d. sample set, in this paper
we estimate the R\'enyi-$\alpha$ divergence using two i.i.d. sample sets.
\end{enumerate}

To the best of our knowledge, only very few consistent nonparametric
estimators exist for R\'enyi-$\alpha$ divergences: \citet{poczos11AISTATS}
proposed a $k$-nearest neighbour based estimator and proved the weak
consistency of the estimator, but they did not study the convergence rate of
the estimator.

\citet{Wang-Kulkarni-Verdu2009} provided an estimator for
the $\alpha \to 1$ limit case only, i.e., for the
\acro{KL}-divergence. They did not study the convergence rate either, and
there is also an apparent error in this work; they applied the reverse Fatou
lemma under conditions when it does not hold. This error originates in the work
 \citet{kozachenko87statistical} and can also be found in other works.
Recently, \citet{perez08estimation} has proposed another consistency proof for
this estimator, but it also contains some errors: he applies the strong law of
large numbers under conditions when it does not hold and also assumes that
convergence in probability implies almost sure convergence.

\citet{hero02alpha,hero2002aes} also investigated the
R\'enyi divergence estimation problem but assumed that one of the two
density functions is known. \citet{gupta12parametric} developed
algorithms for estimating the Shannon entropy and the \acro{KL}
divergence for certain parametric
families.

Recently, \citet{nguyen10estimating}
developed methods for estimating $f$-divergences using their
variational characterization properties. They estimate the likelihood
ratio of the two underlying densities and plug that into the
divergence formulas. This approach involves solving a convex
minimization problem over an infinite-dimensional function space. For
certain function classes defined by reproducing kernel Hilbert spaces
(\acro{RKHS}), however, they were able to reduce the computational
load from solving infinite-dimensional problems to solving
$n$-dimensional problems, where $n$ denotes the sample size. When $n$
is large, solving these convex problems can still be very demanding.
They studied the convergence rate of the estimator, but did not derive
exponential concentration bounds for the estimator.

\citet{2010arXiv1012.4188S,Laurent96Efficient,birge95estimation} studied the
estimation of non-linear functionals of density. They, however, did not study
the R\'enyi divergence estimation and did not derive exponential concentration
bounds either. Using ensemble estimators, \citet{sricharan12ensemble} derived
fast rates for entropy estimation, but they did not investigate the divergence
estimation problem.

\citet{Leonenko-Pronzato-Savani2008} and \citet{goria05new}
considered Shannon and R\'enyi-$\alpha$ entropy estimation from a
single sample.\footnote{The original presentations of these works
contained some errors; \citet{leonenko10correction} provide
corrections for some of these theorems.} In this work, we study
divergence estimators using two independent samples.   Recently,
\citet{pal10estimation} 
proposed a method for consistent R\'enyi information estimation, but
this estimator also uses one sample only and cannot be used for estimating
R\'enyi divergences.

Further information and useful reviews of several
different divergences can be found, e.g.,
in \citet{villmann10mathematical}. 